\documentclass[aps,prx,groupedaddress,reprint]{revtex4-1}
\usepackage{graphicx}
\usepackage{dcolumn}
\usepackage{bm}
\usepackage{amsmath}

\begin{document}

\title{Phase Random Walk Trace in High-order Coherence of Two First-order Incoherent Sources}

\author{Peilong Hong}
\affiliation{The MOE Key Laboratory of Weak Light Nonlinear Photonics and School of Physics, Nankai University, Tianjin 300457, China}

\author{Guoquan Zhang}
\email{zhanggq@nankai.edu.cn}
\affiliation{The MOE Key Laboratory of Weak Light Nonlinear Photonics and School of Physics, Nankai University, Tianjin 300457, China}

\date{\today}

\begin{abstract}

High-order coherence effects between two first-order incoherent sources with fully independent phases have been well studied in the literature, which shows interference fringes with respect to the position separations among different space points. Here we show that this is not the whole story, and find that the high-order coherence effects depend on the mode of the phase random walk of the first-order incoherent sources, which can be controlled artificially and represented geometrically by vectorial polygons. Interestingly, by scanning the detectors along the same direction with the position separations between them kept constant, a set of high-order coherence fringes, which fingerprint the phase random walk of the first-order incoherent sources, can be observed. Our results show that it is possible to control the high-order coherence of two first-order incoherent sources, which could have important practical applications such as superhigh resolution optical lithography.

\end{abstract}


\maketitle

\section{Introduction}
Optical high-order coherence effect was first reported by Hanbury Brown and Twiss (HBT) in 1956, where interference of a thermal source consisting of many first-order incoherent point sources forms the bunching effect in the far field plane~\cite{HBT}. From then on, lots of attentions have been attracted to the field of optical high-order coherence, leading to discovery of many intriguing interference effects~\cite{LOUDON00}. In the field of optical coherence, one of the most fundamentally important subjects is the two beam interference, which provides the fundamental issues of interference effects such as those observed in various interferometers and during the light propagation in free space, and it can be historically traced back to Yang's double-slit interference, where the phase difference between two beams play the key role in the first-order coherence. For high-order coherence, this is also true when one considers coherence of the light field superposed by two first-order incoherent beams from two first-order incoherent sources~\cite{mandel1965coherence,mandel1983photon,paul1986interference,ou1988quantum,klyshko1994quantum,MANDELRMP99,agafonov2008high}, which offers the basic understanding of many other high-order coherence effects~\cite{hong1987measurement,franson1989bell,Jacobson95} and also results in many applications such as ghost imaging~\cite{pittman1995optical,gatti2004ghost,chen2010high,zhou2010third}, subwavelength interference and optical lithography ~\cite{Boto00,Angelo01,Xiong05,jian10,Hong13,Fonseca99} and super-resolving measurements~\cite{hong1987measurement,Oppel2012,Walther04,Mitchell04,Afek10}. 

Now, it is well known that high-order coherence effects between two first-order incoherent sources will show interference fringes with respect to the position separations among different space points, which is usually known as the HBT interference effect. For example, two-photon interference appears as a periodical cosine fringe with respect to the distance between two observation points~\cite{mandel1983photon}. In fact, previous studies on the subject usually presupposed the fully independent random phase of the first-order incoherent sources without taking further consideration on the details of the phase random walk of the sources, which may play an important role on the high-order coherence, just as the role of the phase difference on the first-order coherence of two beams. In this paper, we consider different modes of phase random walks of two first-order incoherent sources, and show that the high-order interference patterns can be controlled through the phase random walk of the first-order incoherent sources. Furthermore, the statistic trace of the phase random walk of the sources can be extracted from the observed high-order interference pattern.

\begin{figure}[b]
\centering
\includegraphics[width=0.45\textwidth]{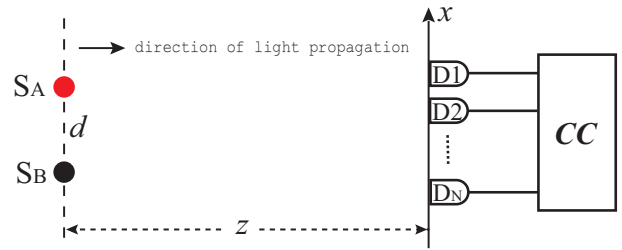}
\caption{General scheme to measure the first- and high-order coherence between two sources $S_A$ and $S_B$. $CC$: Coincidence count. \label{scheme}}
\end{figure}

\section{Brief review}

Figure~\ref{scheme} shows the general scheme to measure the first and high-order coherence effects between two spatially separated sources $S_A$ and $S_B$ by using several single-photon detectors $D_i$ ($i=1, 2, \cdots, N$). Here the distance between $S_A$ and $S_B$ is $d$, and that between the observation plane and the source plane is $z$. If the two sources are coherent to each other, the scheme is essentially a Yang's double slit interference scheme, which will show first-order coherence with typical periodical cosine interference fringes in the far field plane, i.e.,  $I(x) \propto 1+\cos(kdx/z)$, with $k$ being the wave vector and $x$ being the coordinate of the observation point. Such a periodical fringes will be erased if the two sources $S_A$ and $S_B$ are independent or first-order incoherent, which can be realized by introducing a temporally changed initial phase $\varphi$ for source $S_B$, for example, through a phase modulator. However, interference fringes will appear in this case when the second- and higher-order coherence are considered.

Let's first have a brief review on the well discussed situation when the phase $\varphi$ is fully randomly changed, which is equivalent to the case when the phase $\varphi$ is linearly and randomly distributed in the range $[0, 2\pi)$. In this case, the intensity in the observation plane fluctuates randomly with $\varphi$, and the first-order coherence will be erased since the ensemble averaged intensity distribution $I(x) \propto \langle 1+\cos(kdx/z+\varphi)\rangle = 1$, where $\langle \cdots \rangle$  means ensemble average. The second-order correlation function is calculated as

\begin{figure}[b]
\centering
\includegraphics[width=0.42\textwidth]{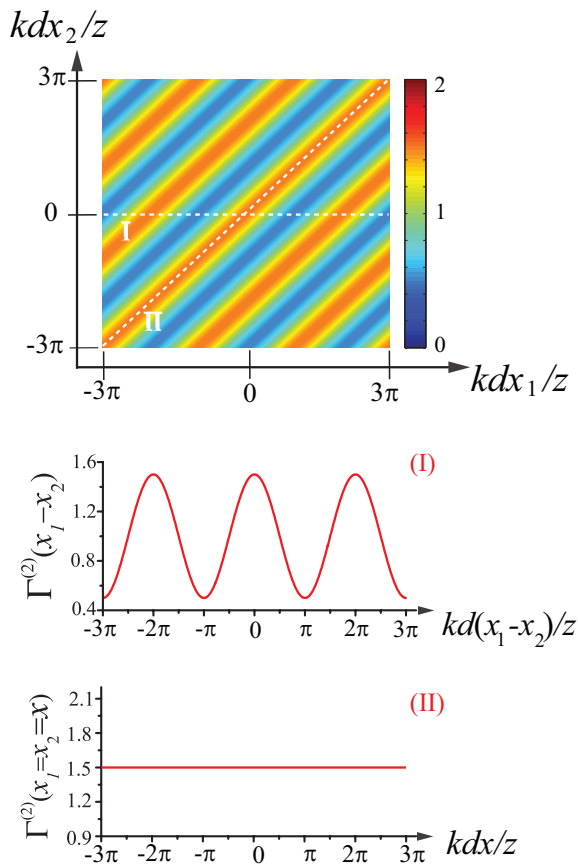}
\caption{Second-order coherence pattern when the phase $\varphi$ changes randomly and linearly within the range $[0, 2\pi)$. \label{case-1}}
\end{figure}

\begin{equation}\label{G2-1}
\begin{split}
\Gamma^{(2)}(x_1, x_2) = & \langle I(x_1,\varphi)I(x_2,\varphi) \rangle \\
\propto&\big\langle \big(1+\cos(kdx_1/z+\varphi)\big)\\
&\times\big(1+\cos(kdx_2/z+\varphi)\big) \big\rangle\\
=&1+0.5\cos(kd(x_1-x_2)/z)
\end{split}
\end{equation}
where $x_1$ and $x_2$ are the coordinates of two space points in the observation plane, respectively. Figure~\ref{case-1} gives
the 2-dimensional coherence pattern with respect to $x_1$ and $x_2$ described by Eq.~(\ref{G2-1}), and the magnitude of the second-order
correlation function $\Gamma^{(2)}(x_1, x_2)$ is represented via pseudocolor. The well-known periodical cosine interference fringes
versus $x_1-x_2$ with a visibility of 50\% could be observed by keeping one detector fixed while scanning the other detector along line I in Fig.~\ref{case-1}~\cite{HBT,mandel1983photon,jian10}. However, if one scans the two detectors together along line II in Fig.~\ref{case-1}, e.g., scans with $x_1 = x_2 = x$, the fringes disappear since $\Gamma^{(2)}(x_1= x_2=x)=constant$. For the third- and higher-order coherence, the interference fringes will appear slightly different, but also with respect to the position separations among different space points~\cite{agafonov2008high,chen2010high,zhou2010third,Oppel2012}.
The above results are based on the assumption that the phase $\varphi$ goes through a fully random walk process, in which the value of one step could be any one within the range $[0, 2\pi)$ and the probability of each step with any value is the same. In fact, there may be other modes of random walk which could also lead to first-order incoherence of two initially coherent sources, and therefore to the appearance of new high-order coherence effects.

\section{2-step random walk}

In a simplest case, let's consider the case when the phase $\varphi$ goes through a 2-step mode of random walk process, in which the values of the two steps are $\{\theta_1, \theta_2\}$ with their respective walking probability $P_1$ and $P_2$, respectively. Considering the requirement of first-order incoherence between two sources, the ensemble averaged intensity in the detection plane should be a constant

\begin{equation}\label{I-2}
\begin{split}
I(x) = &\langle I(x,\varphi)\rangle \\
\propto&1+P_1 \cos(kdx/z+\theta_1)
+P_2 \cos(kdx/z+\theta_2)\\
=&1
\end{split}
\end{equation}
Clearly, there is no single-photon interference when $P_1=P_2=0.5$ and $\theta_2=\theta_1+\pi$. Under this condition, the second-order intensity correlation function can be calculated as
\begin{equation}\label{G2-2}
\begin{split}
\Gamma_{2-step}^{(2)}(x_1, x_2) = & \langle I(x_1,\varphi)I(x_2,\varphi) \rangle \\
\propto &1+\langle \cos(kdx_1/z+\varphi)\cos(kdx_2/z+\varphi) \rangle \\
=&1+0.5\cos(kd(x_1-x_2)/z) \\
&+0.5\cos(kd(x_1+x_2)/z+2\theta_1)
\end{split}
\end{equation}

\begin{figure}[b]
\centering
\includegraphics[width=0.45\textwidth]{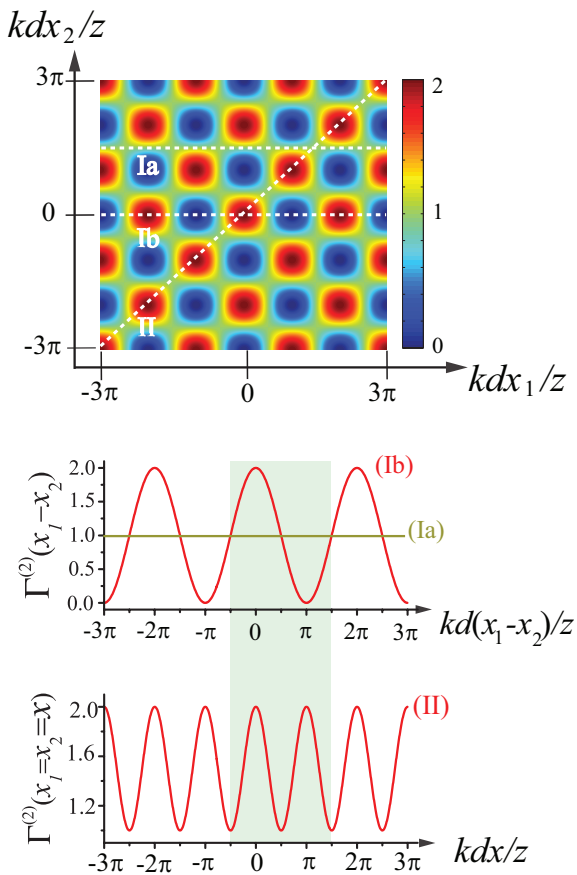}
\caption{Second-order coherence pattern when the phase $\varphi$ goes through a 2-step mode of random walk, in which the values of two steps are $\{0, \pi\}$ and the probability of each walking step is 0.5.\label{case-2}}
\end{figure}

The result shows a second-order coherence pattern completely different from those discussed previously. Besides
the well-known cosine fringes $\cos(kd(x_1-x_2)/z)$ with respect to $(x_1-x_2)$, another kind of cosine fringes
$\cos(kd(x_1+x_2)/z+2\theta_1)$  with respect to  $(x_1+x_2)$ is included, which is similar for any $\theta_1$ but with a phase shift $2\theta_1$. Superposition of these two kinds of interference fringes makes the total interference pattern quite richer as shown in Fig.~\ref{case-2}. One notes that, when keeping one detector spatially fixed while scanning
the other detector, the second-order interference fringes will depend on the position of the fixed detector.
As shown in  Fig.~\ref{case-2}, a constant second-order correlation function is obtained by scanning the detector along line Ia, while scanning the detectors along line Ib gives a periodical cosine fringes with a visibility of 100\%
which is usually thought as a property for two-photon interference with quantum single-photon sources. By fixing
one detector on some other positions and scanning the other detector, one can obtain similar interference fringes
but with different visibility, including the well-known periodical cosine fringes with a visibility of 50\%. Besides, if one sets the two detectors at the same position with  $x_1 = x_2 = x$ and scans them together along line II, the result is no longer a constant but shows subwavelength interference with respect to $x$ as shown in  Fig.~\ref{case-2}, which is helpful to improve the resolution of optical lithography~\cite{Boto00,Angelo01} and reach super resolution phase measurement~\cite{Fonseca99,Oppel2012,Walther04,Mitchell04,Afek10}.

In general, the $N$th-order coherence can be predicted by calculating the $N$th-order intensity correlation function
\begin{equation}\label{GN-21}
\begin{split}
\Gamma_{2-step}^{(N)}&(x_1, x_2, \cdots,x_N) = \langle \prod_{i=1}^{N} I(x_i,\varphi) \rangle \\
\propto &\langle \prod_{i=1}^{N} 2\cos^2((kdx_i/z+\varphi)/2) \rangle \\
=& 0.5\times \prod_{i=1}^{N} 2\cos^2((kdx_i/z+\theta_1)/2) \\
&+0.5\times \prod_{i=1}^{N} 2\cos^2((kdx_i/z+\theta_1+\pi)/2)
\end{split}
\end{equation}
which shows very complicated interference fringes with respect to the position of each space point $x_i$. To clearly show the difference among the high-order interference fringes of the two first-order incoherent sources with different modes of phase random walks, we will consider the $N$th-order correlation function with  $x_1 = x_2 = \cdots = x_N = x$. In this case, the part of traditional HBT-type $N$th-order interference fringes with respect to the position separations between different space points will disappear and only contributes a constant value, therefore, the remnant interference fringes is only related to the mode of the phase random walk of the first-order incoherent sources. The $N$th-order coherence function evolves to

\begin{equation}\label{GN-22}
\begin{split}
\Gamma_{2-step}^{(N)}(x) = & \langle I(x,\varphi)^N \rangle \\
\propto &0.5\times[2\cos^2((kdx/z+\theta_1)/2)]^N \\
&+0.5\times[2\cos^2((kdx/z+\theta_1+\pi)/2)]^N
\end{split}
\end{equation}
which shows a periodical interference fringes with respect to $x$. Two elementary fringes in the form of $P_i\times[2\cos^2((kdx/z+\theta_i)/2)]^N$ are included but with a relative phase shift $\pi$ when we take $x'=kdx/z$ as a variable entity in Eq.~(\ref{GN-22}). It is the superposition of these two elementary fringes that makes the period of the total interference fringes half of that of the single-photon interference when the two sources are coherent to each other. Furthermore, the visibility of the interference fringes grows with $N$.

\section{3-step random walk}
A more complicated case comes when the phase $\varphi$ goes through a 3-step mode of random walk among $\{\theta_1, \theta_2, \theta_3\}$ with a probability of each random walk being $P_1$, $P_2$ and $P_3$, respectively. Again, the requirement that the ensemble average of the intensity is a constant gives rise to

\begin{equation}\label{condition-3}
\begin{split}
P_1 \cos(kdx/z+\theta_1)&+P_2 \cos(kdx/z+\theta_2)\\
&+P_3 \cos(kdx/z+\theta_3)=0
\end{split}
\end{equation}

\begin{figure}[!htb]
\centering
\includegraphics[width=0.35\textwidth]{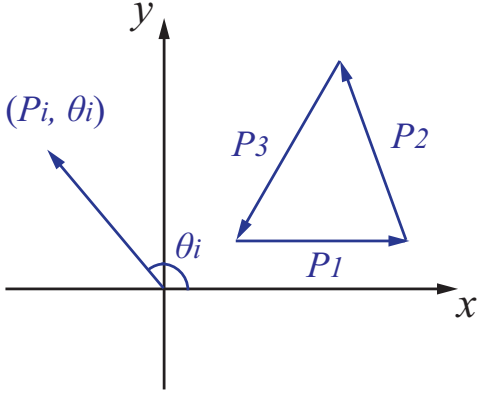}
\caption{ Vector $(P_i, \theta_i)$ represented in the Cartesian coordinate system and the vectorial triangle constructed by three vectors $(P_i, \theta_i)$ $(i=1, 2, 3)$. \label{vector}}
\end{figure}

To find out the condition of Eq. (\ref{condition-3}), we can turn to a more intuitive geometric way, i.e., the vectorial triangle configuration. As shown in Fig.~\ref{vector}, a vectorial triangle is constructed by three vectors $(P_i, \theta_i)$ with the module $P_i$ and the argument $\theta_i$, respectively. Because the sum of the three vectors is zero, it immediately leads to

\begin{equation}\label{M-1}
P_1 \cos(\theta_1)+P_2 \cos(\theta_2)+P_3 \cos(\theta_3)=0
\end{equation}
Note that the equation is also satisfied by rotating each vector  $(P_i, \theta_i)$ with an arbitrary angle $\theta_0$, resulting in the general formula

\begin{equation}\label{M-2}
P_1 \cos(\theta_1+\theta_0)+P_2 \cos(\theta_2+\theta_0)+P_3 \cos(\theta_3+\theta_0)=0
\end{equation}
which is the same as Eq.~(\ref{condition-3}) when $\theta_0 = kdx/z$. Therefore, we can conclude that \emph{if the three vectors $(P_i, \theta_i)$  could construct a vectorial triangle, then the first-order coherence between two sources will be erased}. Under this condition, the $N$th-order coherence can be predicted by calculating the $N$th-order intensity correlation function $\Gamma_{3-step}^{(N)}(x_1, x_2, \cdots,x_N)=\langle \prod_{i=1}^{N} I(x_i,\varphi) \rangle $, which shows pretty rich interference pattern with respect to the space points $x_i$ independently.

Again, for simplicity and clarity, we only consider the $N$th-order correlation function with $x_1=x_2=\cdots=x_N=x$ as

\begin{figure}[!htb]
\centering
\includegraphics[width=0.48\textwidth]{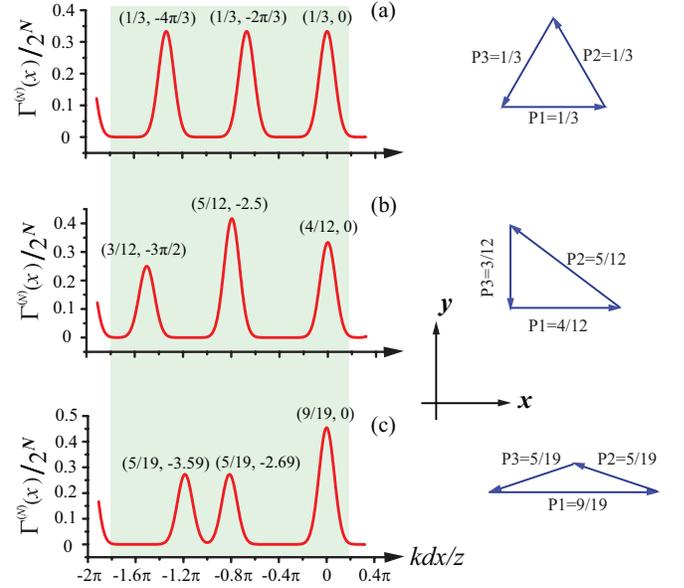}
\caption{Fingerprints of the phase random walks or the vectorial triangles in the $N$th-order coherence fringes with $N=50$ and $x_1=x_2=\cdots=x_N=x$, where the phase $\varphi$ goes through three different 3-step modes of random walks $(1/3, -4\pi/3)$, $(1/3, -2\pi/3)$, $(1/3, 0)$ for (a), $(3/12, -3\pi/2)$, $(5/12, -2.50)$, $(4/12, 0)$ for (b), and $(5/19, -3.59)$, $(5/19, -2.69)$, $(9/19, 0)$ for (c), respectively.\label{case-3}}
\end{figure}

\begin{equation}\label{GN-3}
\begin{split}
\Gamma_{3-step}^{(N)}(x) = & \langle I(x,\varphi)^N \rangle \\
\propto &P_1\times[2\cos^2((kdx/z+\theta_1)/2)]^N \\
&+P_2\times[2\cos^2((kdx/z+\theta_2)/2)]^N\\
&+P_3\times[2\cos^2((kdx/z+\theta_3)/2)]^N
\end{split}
\end{equation}
which shows a similar interference pattern as $\Gamma_{2-step }^{(N)}(x)$ for the case of 2-step mode of random walk expressed by Eq.~(\ref{GN-22}). However, there are three elementary fringes $P_i\times[2\cos^2((kdx/z+\theta_i)/2)]^N$ in Eq.~(\ref{GN-3}), and each of them is characterized with a probability amplitude $P_i$ and a phase shift $\theta_i$, which are the fingerprint of the phase random walk or the vectorial triangle. As shown in Fig.~\ref{case-3}, three different 3-step random walks represented by three different vectorial triangles are selected to erase the first-order coherence between $S_A$ and $S_B$, leading to three different kinds of interference fringes, and the fingerprint $(P_i, \theta_i)$ of the selected phase random walks or the vectorial triangles can be extracted easily from the respective elementary fringes in one period of the interference fringes. This result shows that not only the spatial information of the two sources but also the information of phase random walk to erase the first-order coherence between the two sources can be extracted from the observed interference pattern in the far field plane.

\section{General case and discussions}
If one considers other modes with more steps for phase $\varphi$ to randomly walk, the requirement of first-order incoherence can be satisfied by introducing various vectorial polygons, such as different vectorial quadrilaterals, pentagons, etc, and the high-order interference patterns could be very different for each mode, depending on the shape of the selected vectorial polygon. In general, the $N$th-order intensity correlation function associated with a $M$-side vectorial polygon is $\Gamma_{M-step }^{(N)}(x_1, x_2, \cdots,x_N)=\langle \prod_{i=1}^{N} I(x_i,\varphi) \rangle$. By taking the ensemble average over the $M$-step phase random walk and with $x_1=x_2=\cdots=x_N=x$, the $N$th-order intensity correlation function evolves to

\begin{equation}\label{GN-M}
\begin{split}
\Gamma_{M-step }^{(N)}(x) = & \langle I(x,\varphi)^N \rangle \\
\propto & \sum_{i=1}^M P_i\times[2\cos^2((kdx/z+\theta_i)/2)]^N
\end{split}
\end{equation}
which shows a periodical interference pattern with $M$ elementary fringes, and the $i$th elementary fringe is characterized by a probability amplitude $P_i$ and a phase shift $\theta_i$. Again, one sees that the information of the vectorial polygon or the phase random walk can be extracted from the $M$ different elementary fringes in a period. Specifically, when the vectorial polygon is an equilateral one, every two neighboring elementary fringes is then of the same phase shift $2\pi/M$, resulting in an effective period of the interference fringes reduced by a factor of $M$. Clearly, such an effective subwavelength interference effect is quite different from that relying on the photonic de Broglie wave~\cite{Jacobson95,Fonseca99}, which can achieve a superresolution by a factor of $N$ by performing $N$-photon measurement. In contrast, the reduced effective period of the interference fringes here does not depend on the order of the correlation measurement $N$ but relies on the step $M$ of the phase random walk.

In the geometric viewpoint, for the $N$th-order coherence of two fully independent point sources $S_A$ and $S_B$, which is of an ideal infinite-step mode of phase random walk distributed linearly and randomly within the range $[0, 2\pi)$, the corresponding vectorial polygon would actually approach to a circle. Besides, the condition for the $2$-step mode is equivalent to find two vectors with their summation equal to zero. Therefore, one can conclude that the only possible solution for $2$-step mode is $P_1=P_2=0.5$ and  $\theta_2=\theta_1+\pi$ since the two vectors have to be of the same length but of opposite direction.

\section{Conclusion}
In summary, we show that, besides the traditional ideal infinite-step mode of phase random walk, there are many other modes of phase random walk of two first-order incoherent sources, which can lead to various high-order interference patterns. Each mode of phase random walk is geometrically represented by a closed specific vectorial polygon, and the fingerprint of the vectorial polygon or the phase random walk is revealed in the $N$th-order interference fringes with $x_1=x_2=\cdots=x_N=x$. This means that it is possible to get the concrete phase shift and the corresponding probability of the phase random walk by measuring the $N$th-order correlation function of two first-order incoherent sources. These results deepen our basic understanding of the $N$th-order coherence of light and may lead to novel optical high-order coherence effects and applications such as super-resolving optical lithography.

\section*{acknowledgments}
This work is financially supported by the 973 program (Grant No. 2013CB328702), the NSFC (Grant No. 11174153), and the 111 project (Grant No. B07013).

\end{document}